\documentclass[twocolumn]{aastex62}

\usepackage{soul}
\usepackage{amsfonts,amsbsy}
\usepackage{mathrsfs}
\usepackage{bm} 

\usepackage{graphicx} 
\usepackage{dcolumn} 

\usepackage{aas_macros}

\usepackage{xcolor}
\usepackage{color}

\usepackage{url}
\usepackage{multirow}

\usepackage{ulem}
\usepackage{enumerate}
\usepackage{textcmds}
\usepackage{mathtools}
\usepackage{booktabs}
\usepackage{tabularx}
\usepackage{lineno}
\usepackage{subfigure}
\usepackage{graphicx}

\newcolumntype{p}{D{,}{\pm}{-1}}

\renewcommand{\figurename}{Fig.}
\renewcommand{\tablename}{Tab.}

\renewcommand{\arraystretch}{1.5}

\begin{document}

\title[]{Constraining the cosmic ray propagation halo thickness using Fermi-LAT observations of high-latitude clouds}


\correspondingauthor{Bing-Qiang Qiao, Yi-Qing Guo}
\email{qiaobq@ihep.ac.cn, guoyq@ihep.ac.cn}

\author{Yu-Hua Yao}
\affiliation{College of Physics, Sichuan University, Chengdu 610064, P.R. China
}
\affiliation{Key Laboratory of Particle Astrophysics,
Institute of High Energy Physics, Chinese Academy of Sciences, Beijing 100049, China
}
\author{Bing-Qiang Qiao}
\affiliation{Key Laboratory of Particle Astrophysics,
Institute of High Energy Physics, Chinese Academy of Sciences, Beijing 100049, China
}

\author{Wei Liu}
\affiliation{Key Laboratory of Particle Astrophysics,
Institute of High Energy Physics, Chinese Academy of Sciences, Beijing 100049, China
}

\author{Qiang Yuan}
\affiliation{
Key Laboratory of Dark Matter and Space Astronomy, Purple Mountain Observatory, Chinese Academy of Sciences, Nanjing 210008, China
}
\affiliation{
School of Astronomy and Space Science, University of Science and Technology of China, Hefei 230026, China
}

\author{Hong-Bo Hu}
\affiliation{Key Laboratory of Particle Astrophysics,
Institute of High Energy Physics, Chinese Academy of Sciences, Beijing 100049, China
}
\affiliation{University of Chinese Academy of Sciences, 19 A Yuquan Rd, Shijingshan District, Beijing 100049, P.R.China}

\author{Xiao-Jun Bi}
\affiliation{Key Laboratory of Particle Astrophysics,
Institute of High Energy Physics, Chinese Academy of Sciences, Beijing 100049, China
}
\affiliation{University of Chinese Academy of Sciences, 19 A Yuquan Rd, Shijingshan District, Beijing 100049, P.R.China}

\author{Chao-Wen Yang}
\affiliation{College of Physics, Sichuan University, Chengdu 610064, P.R. China
}

\author{Yi-Qing Guo}
\affiliation{Key Laboratory of Particle Astrophysics,
Institute of High Energy Physics, Chinese Academy of Sciences, Beijing 100049, China
}
\affiliation{University of Chinese Academy of Sciences, 19 A Yuquan Rd, Shijingshan District, Beijing 100049, P.R.China}

\begin{abstract}
  As a basic characteristic of cosmic ray (CR) propagation, the diffusive halo can advance our understanding of many CR-related studies and indirect dark matter. The method to derive the halo size usually has degeneracy problems thus affected by large uncertainties. The diffusion gamma ray from high-latitude clouds might shed light on the halo size independently. Since the spatially dependent propagation (SDP) model has a better agreement with the observed CRs, compared with conventional propagation model, in this work, we investigate the halo thickness based on SDP model with Fermi-LAT $\rm\gamma$-ray observation on the high- and intermediate-velocity clouds. As a result, in order not to exceed the relative $\gamma$-ray emissivity in the high-latitude clouds, halo thickness should be in the range of $\rm ~3.3\sim9~ kpc$. Moreover, the spatial morphology of $\rm\gamma$-rays estimated based on SDP model under different values of halo thickness are distinctive, which provides us a tool to determine the halo size. We hope that our model could be tested and tuned by multi-wavelength observations in the future.
\end{abstract}

\section{Introduction}
\label{sec:intro}
\par
 The galactic halo model was proposed in 1964 \citep{1964ocr..book.....G} to describe Galactic cosmic rays (CRs) propagation. It assumes that CRs are produced by sources located in the thin Galactic disc and then diffuse by scattering off random magnetic fluctuations in a low-density confinement region (“halo”: half the distance of the boundary measured from Galactic plane in the perpendicular direction is $\rm z_h$) extending well beyond the gaseous disc \citep{2015ARA&A..53..199G}. The theoretical explanation of the formation of CR halo includes the turbulent cascade of MHD waves \citep{2018PhRvL.121b1102E}, an increase of the Alfvén velocity with height \citep{2020ApJ...903..135D}. As a basic characteristic of CR propagation, the halo size is intrinsically connected to many CR-related studies, such as the energy spectrum of CRs \citep{2020ApJ...903..135D},  diffuse $\gamma$-ray emission foreground \citep{2010PhRvL.104j1101A}, indirect dark matter or exotic (astro-)physics searches \citep{2014PhRvD..90h1301L,2016PhRvD..94l3019K,2020A&A...639A..74W}. More about the halo size see \cite{2020ApJ...903..135D}.


\par

  The halo size can be independently constrained with the probe of CRs, secondary positrons, radio emission, and $\gamma$-rays. It is well known that the height of the galactic halo and the normalization of diffusion present a large degeneracy, which is tuned by the secondary-to-primary ratio such as B/C \citep{2016PhRvD..94l3019K,2018JCAP...07..051G,2019SCPMA..6249511Y,2020ApJ...903..135D,2020JCAP...11..027Y}. The most widely used probe to solve the degeneracy is "CR clocks", such as radioactive isotopes as, e.g., $\rm {}^{10}Be/{}^{9}Be$ \citep{2016PhRvD..94l3019K}. Because they are very sensitive to the processes occurring in the halo \citep{2002A&A...381..539D,2015ARA&A..53..199G,2007ARNPS..57..285S}. Nevertheless past measurements of isotopic flux ratio in CR are scarce, limited to low energy and affected by large uncertainties \citep{2018JCAP...07..051G, 2015PhRvC..92d5808T}. Alternatively, elemental ratios (e.g. Be/B, Al/Mg) are used to induce constrains on halo size. With published HEAO3 data, a realistic Monte Carlo diffusion model for the propagation of cosmic rays requires the halo size $\rm 2-3~kpc$ \citep{1998ApJ...506..335W}. With AMS-02 data, a best-fit value of the ratio Be/B $\rm z_h\sim7~kpc$ and a lower limit $\rm z_h\geq 5~kpc$ are found in \citep{2020PhRvD.101b3013E}, but in the work of \citep{2020A&A...639A..74W} estimation with $\rm {}^{10}Be/{}^{9}Be$ and Be/B shows a preference for $\rm z_h\sim4-5~kpc$. As an independent probe, low-energy secondary CR positrons allow us to place a lower bound on the halo size (usually at about $\rm3-4~kpc$), assuming the B/C-reduced degeneracy \citep{2014PhRvD..90h1301L,2017A&A...605A..17B,2017PhRvD..95h3007Y,2018JCAP...01..055R,2020A&A...639A..74W}.
  Experiencing energy losses in the diffusion halo which limits the distance from which positrons reach the earth, they are not very sensitive to the boundaries of the diffusion halo, but rather to the diffusion coefficient. The synchrotron emission from CR leptons in the MHz to GHz radio band also provide information about the magnetised halo height \citep{2013JCAP...03..036D,2013MNRAS.436.2127O,2018JCAP...07..063B,2020A&A...639A..74W}. They are somewhat less sensitive to large values of the halo size, although upper limits in the range of 10-15 kpc have been derived. In addiation, independent halo-thickness constraints can also be derived from diffuse $\rm\gamma$-ray \citep{1977ApJ...217..843S,2000ApJ...537..763S}.

\par
  Unlike charged CRs, the propagation of $\rm\gamma$-ray is mostly unaffected by the interstellar medium (ISM) and Galactic magnetic fields (GMFs), and therefore the emission retain information on the morphology of the emission region. Thanks to the Large Area Telescope (LAT), on board the Fermi $\rm\gamma$-ray observatory \citep{2009ApJ...697.1071A}, provides high quality $\rm\gamma$-ray of the all-sky from 30 MeV to beyond a few hundred GeV \citep{2012ApJ...761...91A}. A global analysis of the diffuse $\rm\gamma$-ray emission measured by the Fermi-LAT satellite actually require a larger halo size than usually assumed and $\rm z_h$ up to $\rm \sim10~kpc$ \citep{2011ApJ...726...81A,2012ApJ...750....3A}. Actually, the $\rm\gamma$-ray emission of large-scale regions is not as sensitive as that from small-size location to the spatial morphology of CRs because the former is an integral effect that is likely to eliminate traces. Thus, the $\gamma$-ray emissivity in peculiar regions with small scale, particularly in the direction of perpendicular to the Galactic plane, have been employed to fulfill this task. The observations with high-velocity clouds (HVCs) and intermediate-velocity clouds (IVCs) set constraints of the halo thickness to less than $\rm 6~kpc$ \citep{2015ApJ...807..161T}.
  \par


  Above studies of halo size with $\gamma$-rays are based on the conventional propagation (CP) model. This model has simple geometry which reflects, however, the most essential features of the real system. It is assumed that the system has the shape of a cylinder with a radius R and a half height $\rm z_h$. Instead of sole propagation halo, this work adopts the spatially dependent propagation (SDP) frame. The diffusion volume in the SDP model is divided into two regions as inner halo (IH: $\rm |z| < \xi z_h$ ) and outer halo (OH: $\rm |z| > \xi z_h$). The size of the IH region is represented by half thickness $\rm \xi z_h$, whereas the OH region’s is $\rm (1 - \xi)z_h$. More description about SDP model please see \citep{2016ChPhC..40a5101J,2016ApJ...819...54G,2018PhRvD..97f3008G,2018ApJ...869..176L,2019JCAP...10..010L}. Compared with the CP model, the SDP one has special spatial-morphology and works well to reproduce the abnormalities of CRs \citep{2018PhRvD..97f3008G,2018ApJ...869..176L}. Besides, we expect the SDP model predictions of spatial distribution and energy spectrum are specific with varying the halo size. Therefore it is necessary to reboot the study of the halo thickness based on SDP model. In fact, our previous work, with CR anisotropy from TeV to PeV energy range, has adopted the SDP model to constrain $\rm z_h$ to be less than $\rm12~kpc$ \citep{2102.13498v1}. In this work, we study the halo size based on the SDP model with Fermi-LAT $\gamma$-ray observation of HVCs and IVCs. The paper is organized in the following way, section 2 describes the research and results, section 3 presents the discussion and outlook. Following secion 4 gives the conclusion.

\section{Research and Results}

  Firstly, we calculate $\gamma$-ray emissivity by employing a set of ready-made  transport configurations with different halo heights. Then, because diffusion properties under different halo size scenarios might be diverse, we tune propagation parameters based on the secondary-to-primary ratios and calculate the $\gamma$-ray emissivity again.

\subsection{The Effects of Varying $\rm z_h$ on $\gamma$-ray Emissivity}
\begin{figure*}[!htb]
\centering
\includegraphics[width=0.45\textwidth]{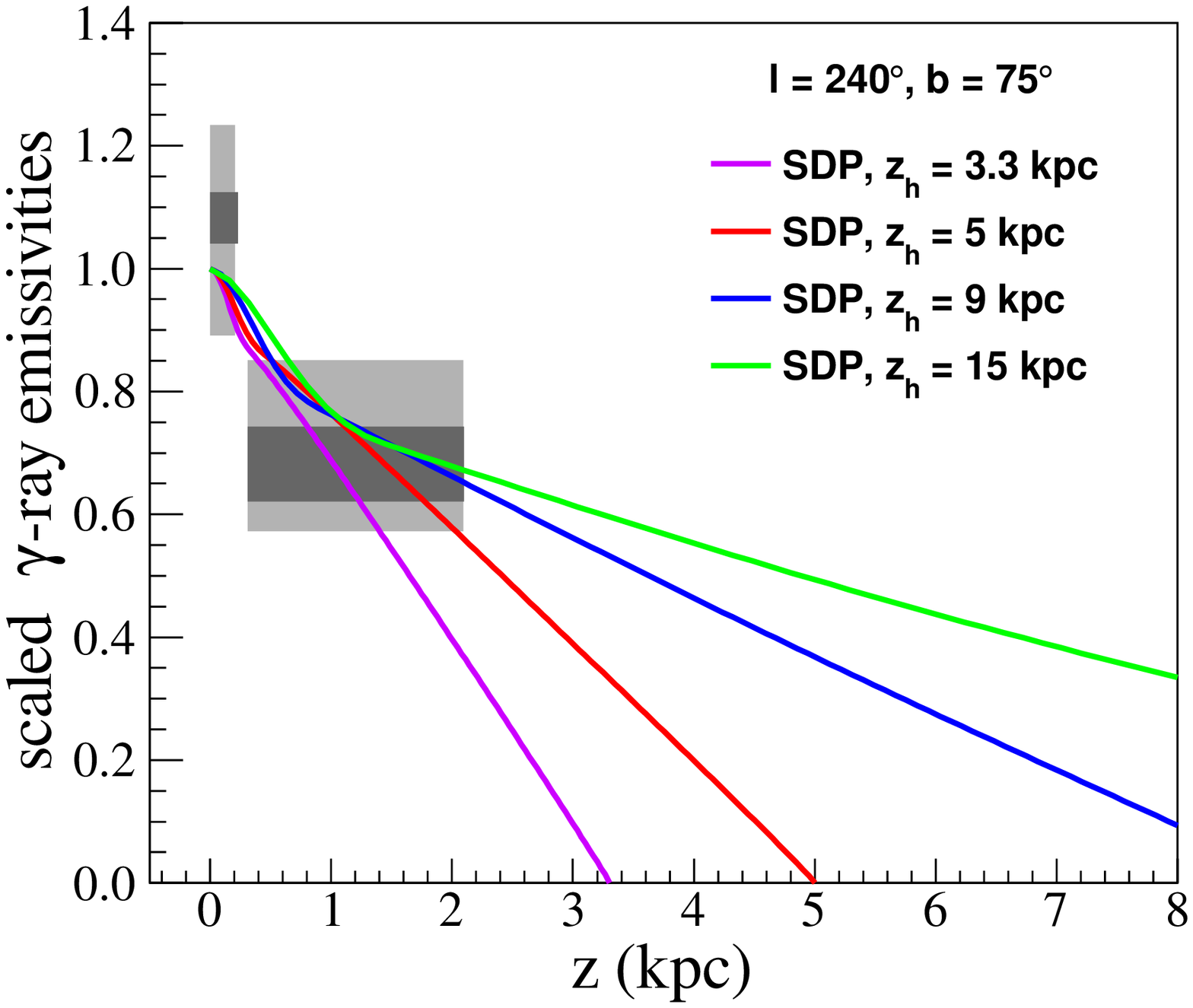}
\includegraphics[width=0.45\textwidth]{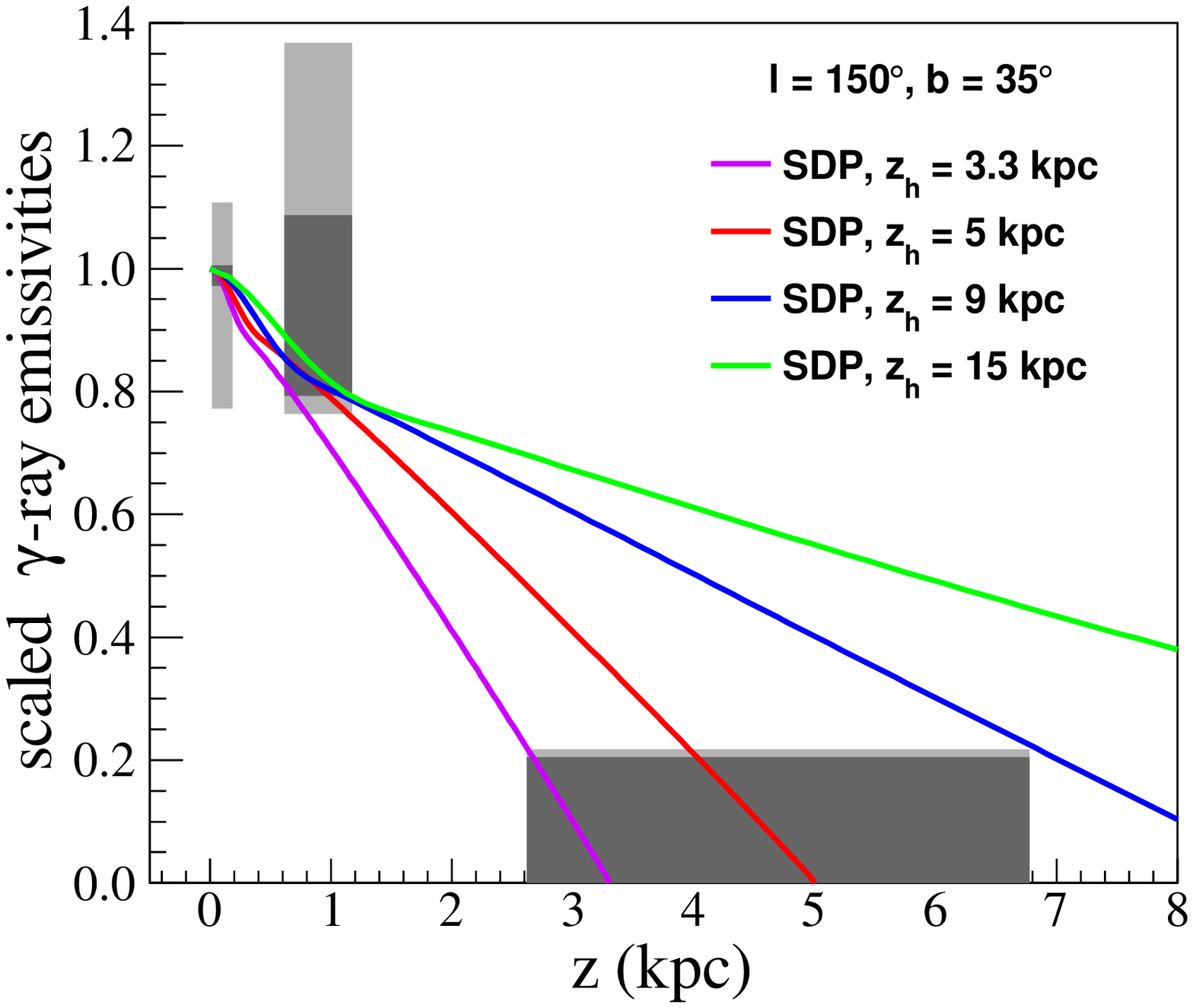}
\caption{
Model predictions of $\gamma$-ray emissivity scaling from 300 MeV to 10 GeV compared with data. A set of ready-made transport configurations under SDP + local source assumption from \citep{2102.13498v1} with different halo heights are employed, detailed parameters are listed in Tab. \ref{tab:para}. The purple, red, blue, and green lines represent SDP model with $\rm z_h =3.3, 5, 9, $ and 15 kpc, respectively. The shaded rectangles are the emissivity scaling factors from Fermi-LAT observation \citep{2015ApJ...807..161T}.
}
\label{fig:halo_size}
\end{figure*}

\begin{figure*}[!htb]
\centering
\includegraphics[width=0.45\textwidth]{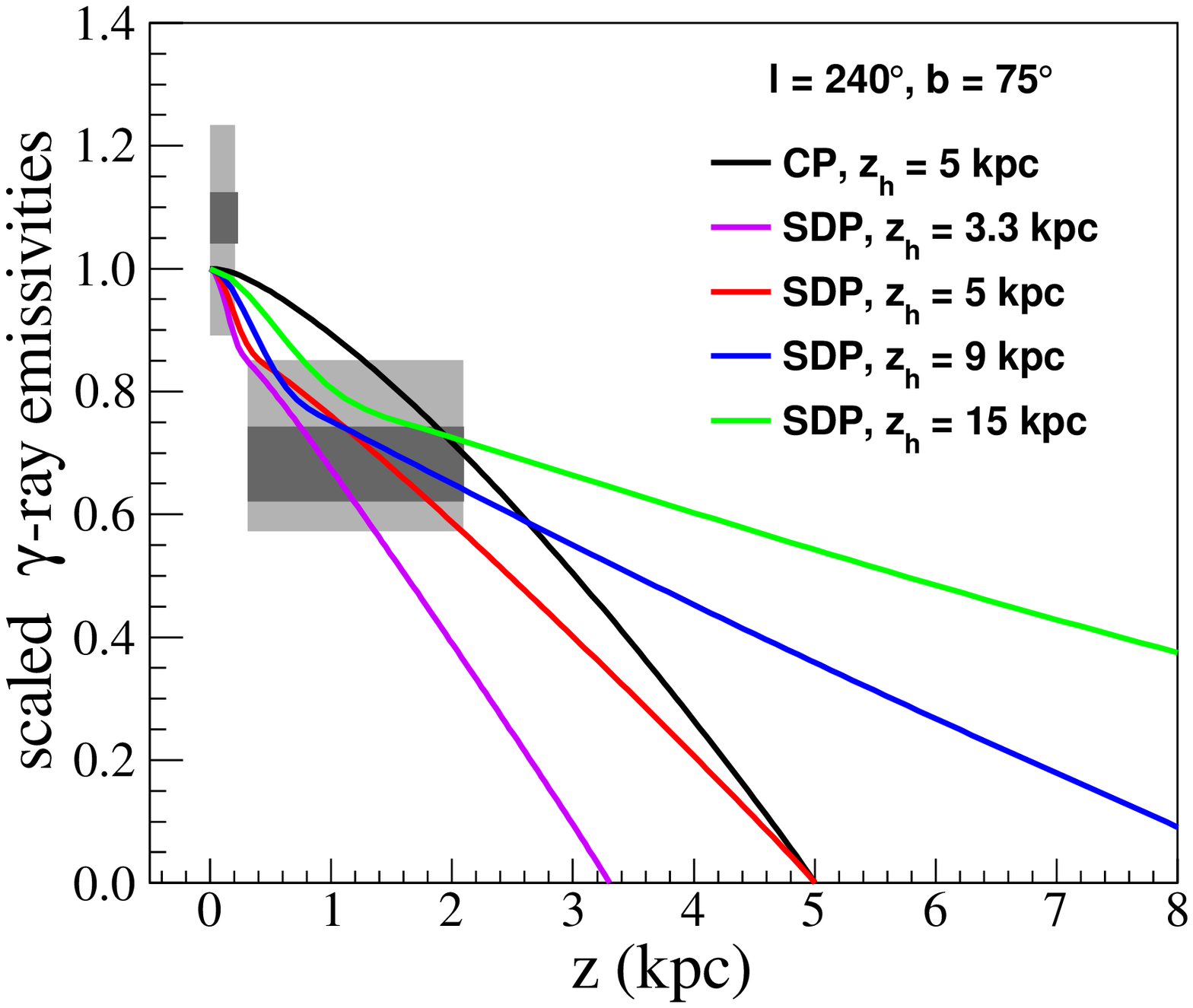}
\includegraphics[width=0.45\textwidth]{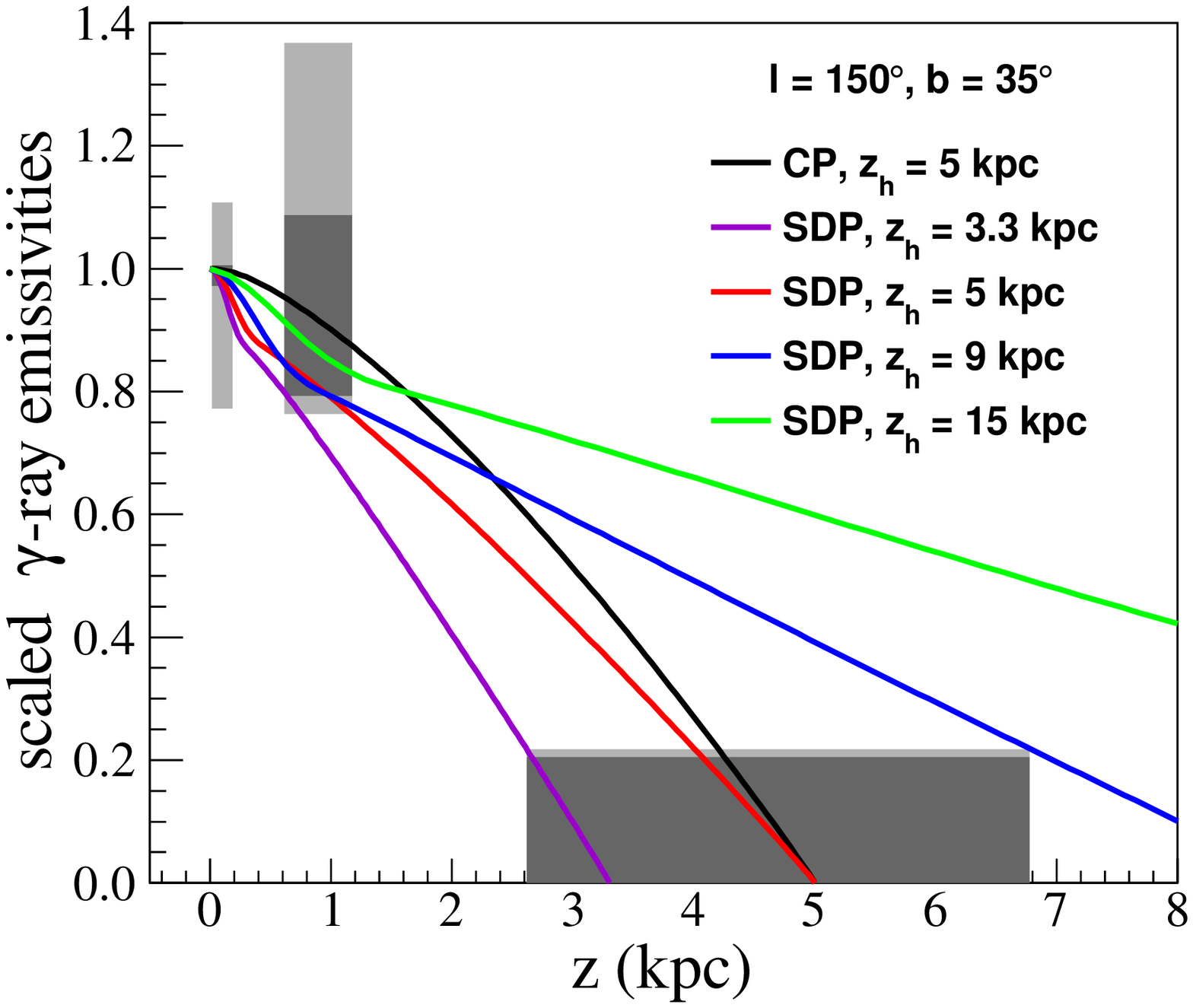}
\caption{
Model predictions of $\gamma$-ray emissivity scaling from 300 MeV to 10 GeV compared with data. Propagation parameters are properly tuned with different halo heights under pure SDP model assumptions and listed in Tab. \ref{tab:para}. The purple, red, blue, and green lines represent SDP model with $\rm z_h =3.3, 5, 9, $ and 15 kpc, respectively. Black lines are from CP model. The shaded rectangles are the emissivity scaling factors from Fermi-LAT observation \citep{2015ApJ...807..161T}.
}
\label{fig:halo_prop}
\end{figure*}

  \begin{figure*}[htb]
\centering
\includegraphics[width=0.45\textwidth]{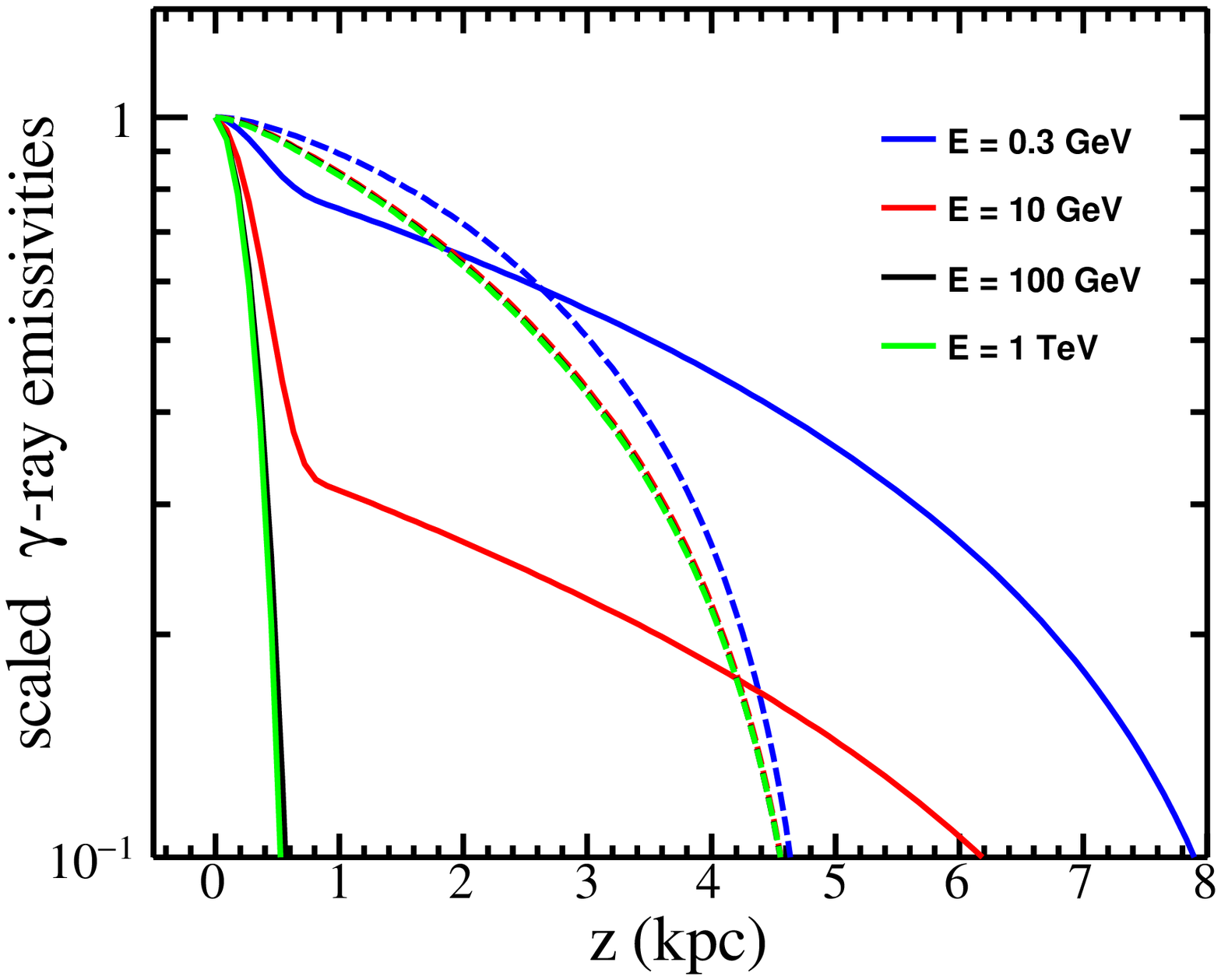}
\includegraphics[width=0.45\textwidth]{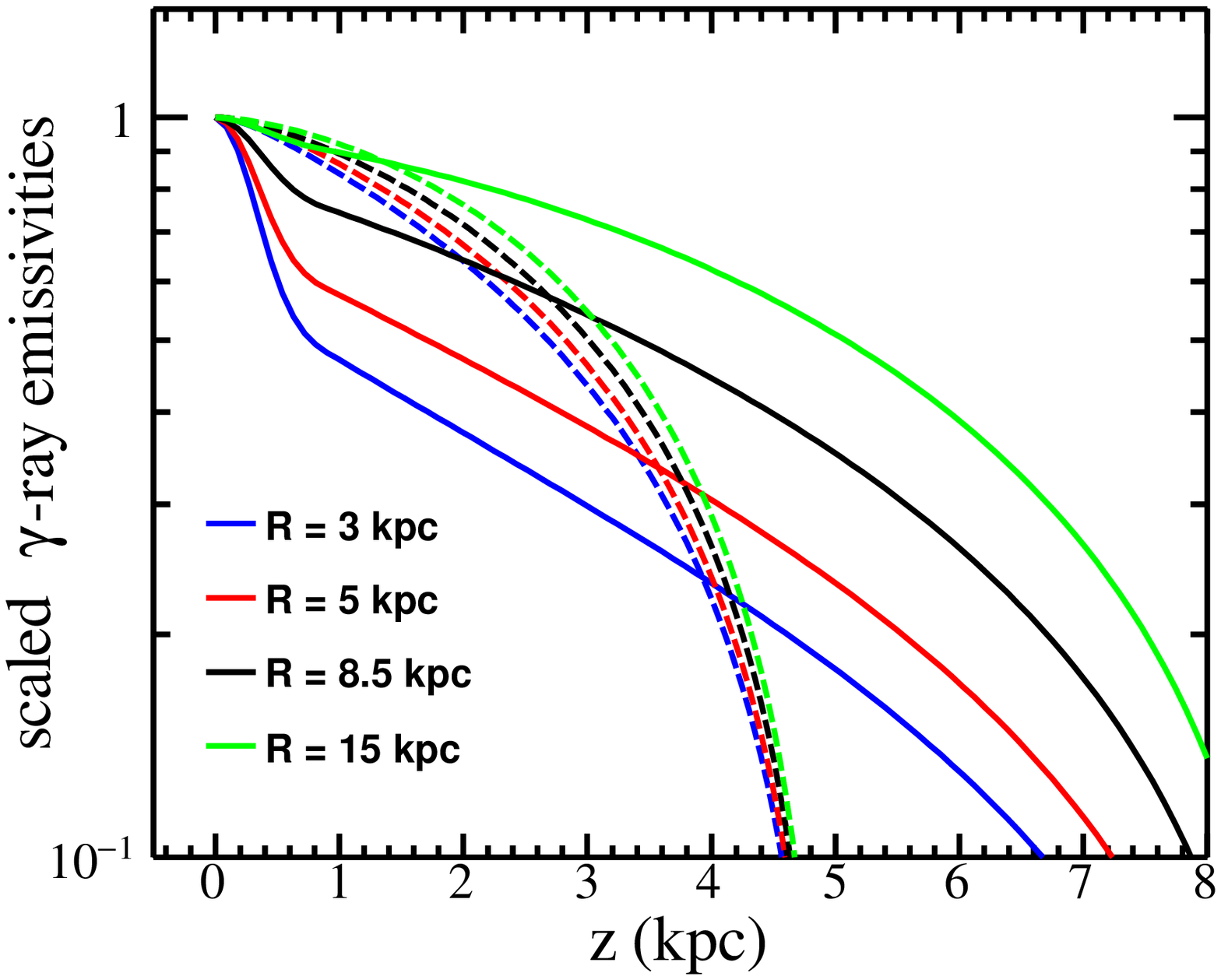}
\caption{
 Model predictions of $\gamma$-ray emissivity from 300 MeV to 10 GeV as a function of vertical scale from the disk. Solid lines represent the contribution from the SDP model and dashed lines represent the contribution from CP model. For SDP model, the set of parameters of $\rm z_h = 9 kpc$ is adopted here.
 (left) The blue, red, black, and green lines represent photon energies of 0.3 GeV, 10 GeV, 10 GeV, and 1 TeV respectively.  (righ) The blue, red, black, and green lines represent the galactocentric radial distances are 3, 5, 8.5, and 15 kpc, respectively.
}
\label{fig:regE}
\end{figure*}

  Propagation parameters of \citep{2102.13498v1} with different halo heights are employed to calculate the $\rm \gamma$-ray emissivity. This set of parameters is under the SDP plus local source assumption, working well to reproduce the anomalies of CR spectrum, CR anisotropy, and so on. Detailed parameters are listed in the Tab. \ref{tab:para}. Fig. \ref{fig:halo_size} shows model curves with the data taken by Fermi-LAT \citep{2015ApJ...807..161T}. The gray rectangles are emissivity scaling factors from Fermi-LAT, which are the ratios of the $\gamma$-ray emissivity ($\gamma$-ray emission rate per hydrogen atom) in each region of HVCs and IVCs over the local emissivity, with measured energies between $\rm 300~MeV$ and $\rm 10~GeV$. The emissivity of local gas is assigned to the range from $\rm z = 0~kpc$ to $\rm z = 0.3~kpc$ (disk). The horizontal widths of the rectangles indicate lower and upper limits on their distances, the vertical height in dark (light gray) corresponding to statistical (total) uncertainties of the emissivity scaling factors. Detailed information about the target regions and their emissivity scaling factors see \citep{2015ApJ...807..161T}. In case of models, we use the CR and gas distribution to calculate the gamma-ray emissivity in the halo and the total gamma-ray flux at the Sun’s position in the Galaxy, for comparison with observations \citep{2015ApJ...807..161T}.


\par
 Fig. \ref{fig:halo_size} shows that there is a broad agreement between models curves and measurements in the region of left panel ($\rm l=240^{\circ}, b=75^{\circ}$), where is an extension of the IV Arch. In the region of right panel ($\rm l=150^{\circ}, b=35^{\circ}$), together with the emissivity of low-latitude IVs, upper limits from the HVC of $\rm z\sim2.6–6.8~kpc$ provide the strongest limit on the value of halo size. In order not to exceed the upper limit, the halo size need to satisfy $\rm 3.3 < z_h < 9~kpc$. It also can be seen that the $\gamma$-ray emissivity in the low latitude (inner halo) are nearly the same whatever the size of the halo. The differences among models increases with latitude thus $\gamma$-ray observations of the mid- and high-latitude clouds would be very valuable.

\par
  Calculation with a set of second-hand fixed propagation parameters and various $\rm z_h$ indicate the relative $\gamma$-ray emissivity changes with the thickness of the halo. However, usually the ratio of halo height and diffusion coefficient is settled, the result of calculation with adjusted diffusion coefficient with the halo height would be more reliable and used to test above statements.

\subsection{The Effects of Varying Propagation Parameters on $\gamma$-ray Emissivity}

 A grid of models are considered with different values of $\rm z_h$. Firstly, fits of the model prediction to the B/C and $\rm {}^{10}Be/{}^{9}Be$ ratios are performed in order to determine the propagation parameters. Then the CR spectra and large-scale all-sky $\rm \gamma$-rays are used for check consistency between predictions from well-turned models and observations. Consequently, the tuned diffusion coefficients are listed in Tab. \ref{tab:para} and the detailed diagram about secondary-to-primary ratios, CR spectra are in the appendix. The $\rm \gamma$-ray emissivity are calculated again and compared with observations. In order for comparison, results from CP model are also illustrated.

\begin{table}[!htb]
\caption{\label{tab:para} Propagation parameters$^\dagger$.}
\begin{ruledtabular}
\begin{tabular}{c|ccccc}
   &$\rm z_h$ & $\rm D_0 (\times 10^{28})$ &  $N_m$ &  $\rm\delta_{0}$ & $v_A$\\
   & \rm kpc & $\rm  cm^{2}~s^{-1}$  &        &                   & \\
\hline
 SDP+local source $^\ddag$&    & 8.75   &   0.39    &  0.65  & 6  \\
\hline
 CP&  5  & 3.72   &   0.24    &  0.46  &22  \\
\hline
    & 3.3 & 3.25 & 0.25&  0.58 & 6\\
   &5 & 5.04   &   0.29    &  0.6   &6 \\
 SDP&9 & 7.25   &   0.37    &  0.65  &6  \\
   &15 & 10.8   &   0.40    &  0.69  &6 \\
\end{tabular}\\
$^\dagger${ n and $\xi$ for SDP model are adopted as 4 and 0.1, respectively.}\\
$^\ddag${This set of parameters is adopted from \citep{2102.13498v1}.}
\end{ruledtabular}
\end{table}

  With propagation parameters listed in Tab. \ref{tab:para}, the $\gamma$-ray emissivity calculation is repeated and presented in Fig. \ref{fig:halo_prop}, which has little difference from model curves in Fig. \ref{fig:halo_size}. This suggests that the diffusion coefficient has a relatively small effect on the distribution of $\rm \gamma$-ray emissivity and the latter could be an independent estimator of the halo height. Because the diffusion coefficient dominates the propagation procedures, the relative flux of $\gamma$-ray between two locations cancel out its impact thus depend heavily on the extend of the halo. In short, the Galactic latitude profile of the relative $\gamma$-ray emissivity are sensitive to the vertical gradient of CR sourcing the emission.

\section{Discussion and Conclusion}

\subsection{Different Energies}

  As the tracer of CRs, we expect the $\gamma$-ray emissivities would vary with energies based on SDP model. Left panel of Fig. \ref{fig:regE} presents $\gamma$-ray emissivities calculated from SDP and CP models assume values of energy between $\rm 0.3~GeV$ and $\rm 1~TeV$. Along with the distance from the disk, the spectra shape estimated from CP model decreases smoothly and the energy has slight influence on them. An increasing galactocentric radius $\rm z$ to $\rm 2~kpc$ produces a decrease of emissivity up to at most $\rm 40\%$ for CP model. However, in view of the SDP model, the shape of emissivity has two segments, corresponding to the transition from IH to OH. Moreover, as energies raise, the scaled fluxes close to the disk decreases more rapidly. The decrease of $\rm 0.3~GeV$ $\gamma$-ray emissivities within $\rm z<3~kpc$ of SDP model is roughly consistent to that of CP model, beyond $\rm 3~kpc$ the SDP model prediction has a slower descent.  As $\gamma$-ray energy increasing to $\rm 10~GeV$,  its emissivities is only about half of intensities of 3 GeV with different values of z. As for even higher energies, the emissivity reduces to less than ten percent within 1 kpc.

\subsection{Different Galactocentric Radii}

  Then, in order to further explore the spatial morphologies of CRs, right panel of Fig.\ref{fig:regE} illustrates the emissivities at different galactocentric radial scale in the disk from 3 kpc to 15 kpc. The emissivity shape of CP model has a limited change due to a constant diffusion coefficient in the whole galaxy. It decreases linearly with z from the disk to the boundary of the halo. But on account of the spatially dependent diffusion coefficient, the emission of the SDP model strongly depends on the galactocentric radius. The larger the galactocentric radius is, the slower the emissivity goes down with z.  When the radius raises to a certain extent, the result from SDP model gradually approaches to the one from CP model. These properties are of great benefits for us to distinguish them with multi-wavelength observations in the future.

\par
 Due to a constant diffusion coefficient, CM model has a gentle spatial-morphology change and limited variation in different energies and galactocentric radiuses. On the contrary, the emission of the SDP model has strong dependence on the galactocentric radius and energies owing to the spatially dependent diffusion coefficient. Note that here the propagation parameters are adopted as $\rm z_{h} = 9~kpc$, these discussions of other parameters should also be valid. In addition, the diffusion volume in the SDP model is divided into two regions, and the diffusive coefficient in these two parts are constant. The halo might have more segments and the real characters of it be more complicated, observations of mid- and high-latitude clouds in the future could test or update our model.

 \par
  In summary, based on the SDP model, we perform the study of the halo thickness. As a result, SDP models with halo heights $\rm z_h\sim3.3-9~ kpc$ are found to provide good fits to the $\rm \gamma$-ray emissivity from high- and intermediate-velocity clouds. The $\rm \gamma$-ray emissivity is a good estimator of halo height and more observations on mid- and high-latitude $\rm \gamma$-ray emissivity could advance our understanding of the specific distribution of CR diffusive halo.

\acknowledgments

This work is supported by the National Key $R\&D$ Program of China grant No. 2018YFA0404202 ,the National Natural Science Foundation of China (Nos. 11635011, 11875264, 11722328, 11851305, U1738205, U2031110).

\appendix
\setcounter{table}{0}   
\setcounter{figure}{0}
\renewcommand{\thetable}{A\arabic{table}}
\renewcommand{\thefigure}{A\arabic{figure}}

\section{Secondary-to-primary ratio}
\begin{figure*}[!htb]
\centering
\includegraphics[width=0.45\textwidth]{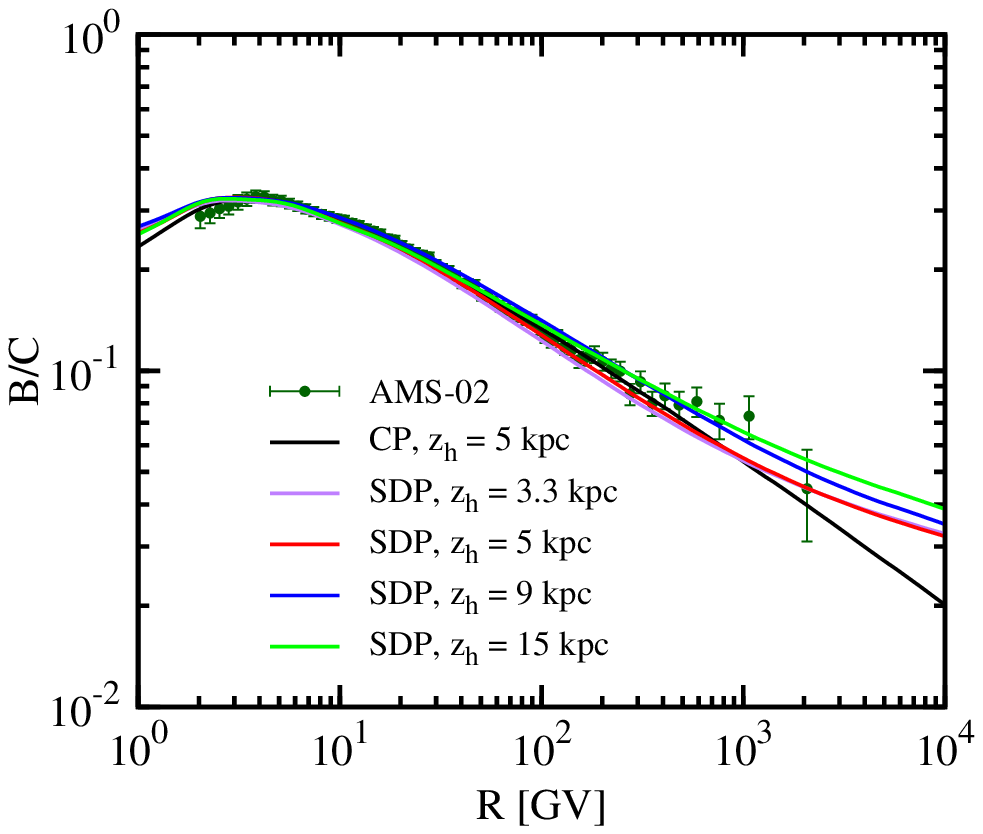}
\includegraphics[width=0.45\textwidth]{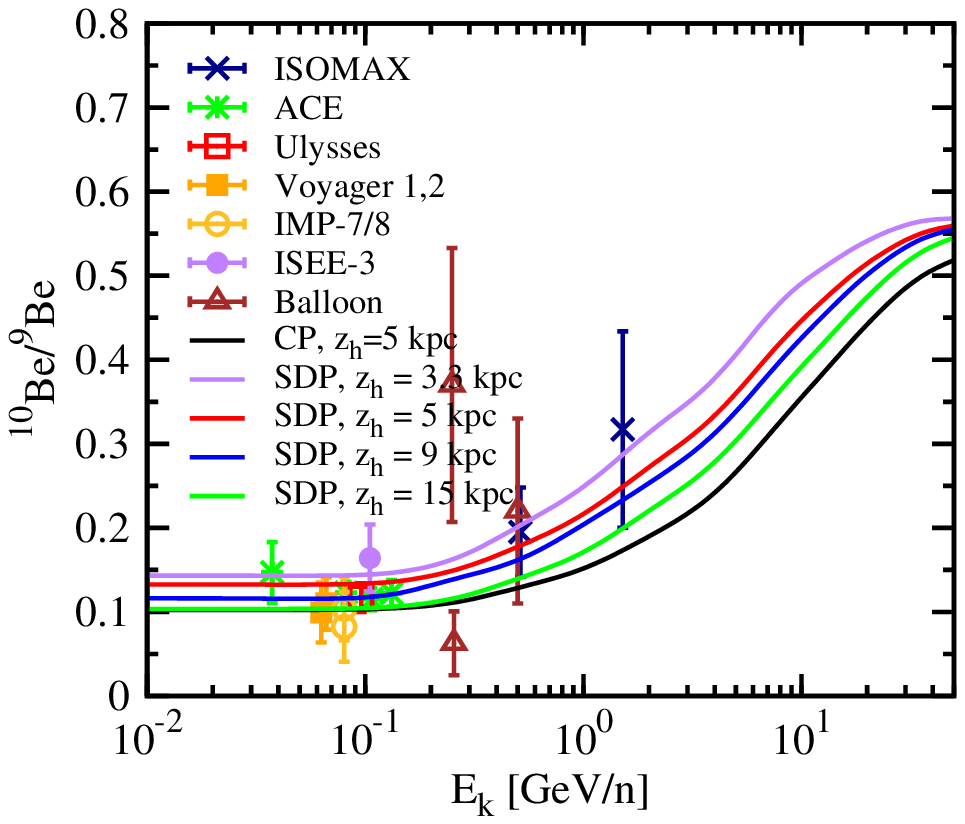}
\caption{
B/C ratio as a function of rigidity (left panel). The B/C data from: AMS02 \citep{2015PhRvL.114q1103A,2017PhRvL.119y1101A}. $\rm Be^{10}/Be^{9}$ ratio as a function of kinetic energy per nucleon (right panel). The $\rm Be^{10}/Be^{9}$ data are from \citep{2015PhRvD..91f3508L} and references therein.
}
\label{fig:sec}
\end{figure*}

  The boron-to-carbon ratio (B/C) has always been considered as the best quantity to study diffusion properties \citep{2005APh....24..146C}. The most precise B/C data have been obtained in the rigidity range $1–10^{3}$ GV by the AMS-02 experiment \citep{2017PhRvL.119y1101A}. In the left panel of Fig. \ref{fig:sec}, along with the experimental data, we plot the theoretical predictions calculated with diffusion model. Lines with different color represent the different values of $\rm z_h$. Besides the secondary-to-primary ratio, the CR isotopic composition of beryllium can also provide unique information on the propagation of CRs in the galaxy.  The ratio of $\rm Be^{10}/Be^{9}$ is shown in the right panel of Fig. \ref{fig:sec}. Consequently, the tuned diffusion coefficients are listed in Tab. \ref{tab:para}. Note that in this work $\rm \xi$ is fixed as 0.1, since slight change of it in fact affect outcomes little.

\section{CR spectra}
\begin{figure*}[!htb]
\centering
\includegraphics[width=0.95\textwidth]{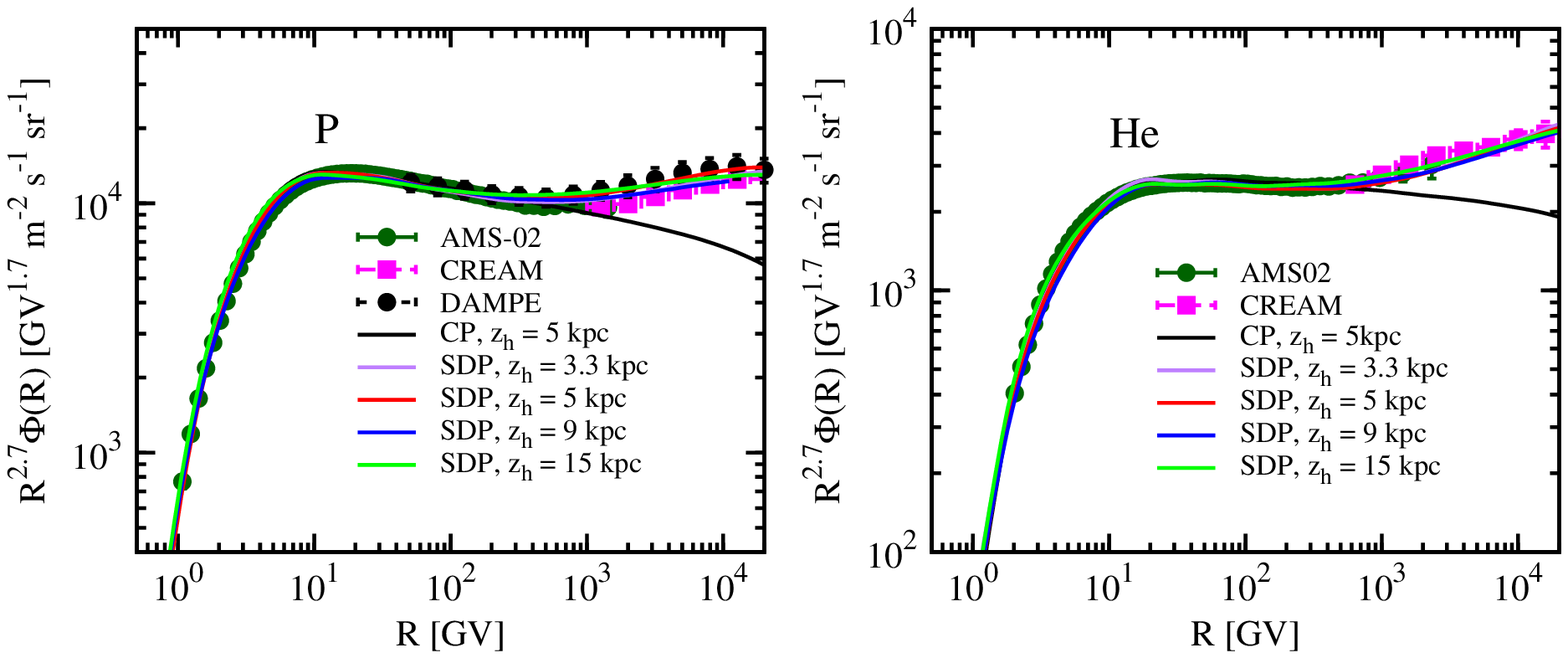}
\caption{
Comparison between model calculations and observations for the primary spectrum of protons (left) and heliums (right). The experiment data come from: AMS-02 \citep{2017PhRvL.119y1101A}, CREAM \citep{2010ApJ...714L..89A}, DAMPE \citep{2019SciA....5.3793A}. Propagation parameters are listed in Tab. \ref{tab:para}.
}
\label{fig:proton}
\end{figure*}

The left of Fig. \ref{fig:proton} shows the proton spectrum, to which we pay particular attention because protons provide the dominant contribution to the diffuse $\gamma$-ray spectra. As we can see that, the spectra from the SDP model of different $\rm~z_h$ match the data well, except the CP model at high energy end. And the same as the helium spectra shows in right panel of Fig. \ref{fig:proton}.

\section{Diffuse $\gamma$-rays}

\begin{figure*}[!htb]
\centering
\includegraphics[width=0.95\textwidth]{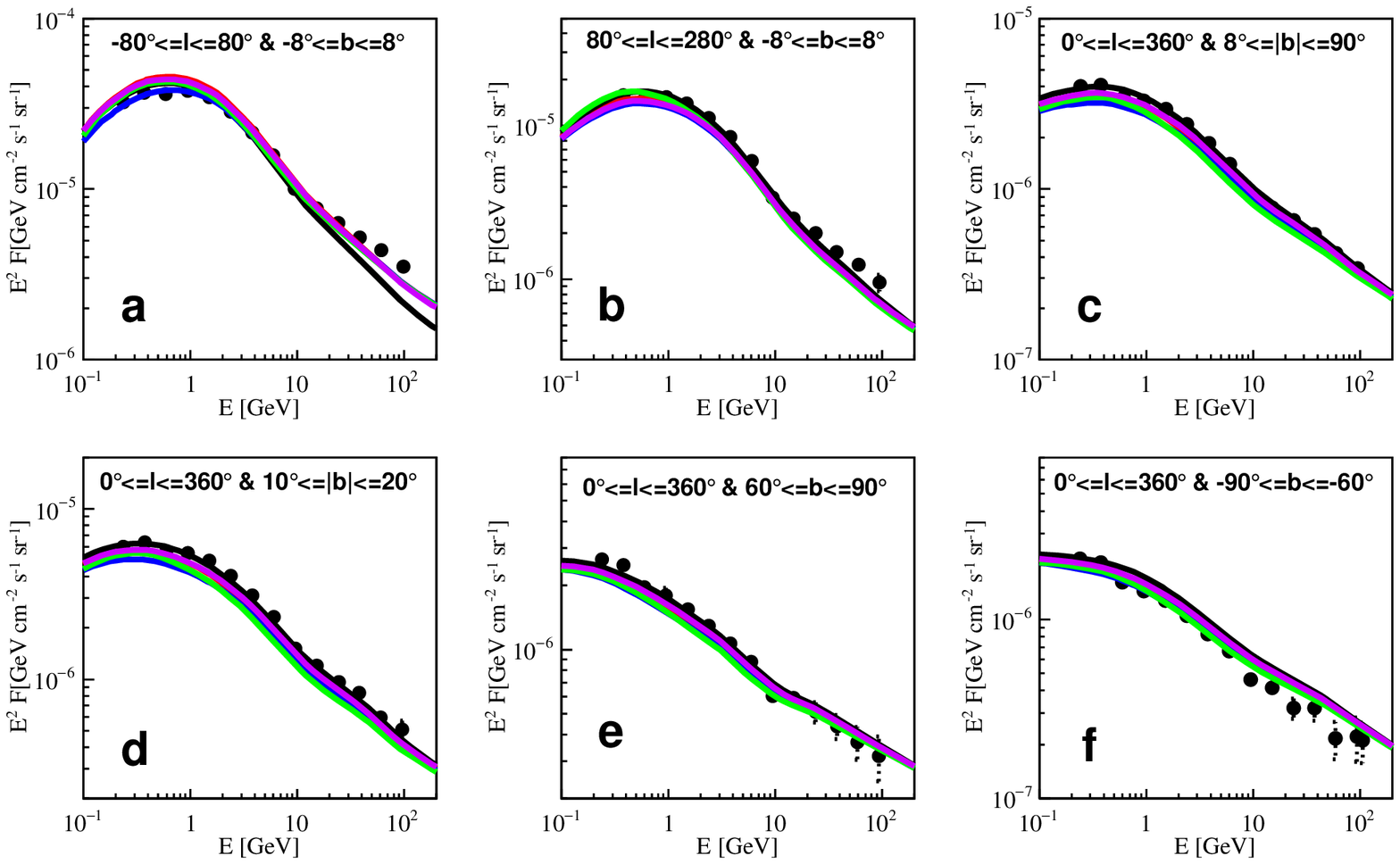}
\caption{
Diffuse $\gamma$-rays at different sky regions. Each plot has several lines, the black one is CP model, purple, red, blue, and green lines represent SDP model at $ z_h = 3.3, 5, 9, $ and 15 kpc, respectively. The data with black points is from Fermi-LAT \citep{2012ApJ...750....3A}. Propagation parameters are listed in Tab. \ref{tab:para}.
}
\label{fig:diffM}
\end{figure*}

Once the parameters of the propagation model have been determined, the predicted $\gamma$-ray maps are compared to the Fermi-LAT data. Fig. \ref{fig:diffM} gives the calculated total $\rm\gamma$-rays spectra by varying the halo size ($\rm z_h$). The purple, red, blue, and green lines present the result from the SDP model at $\rm z_h=3.3, 5, 9, 15~kpc$, respectively, as well as the CP model is plotted as the black line. The model predictions of $\gamma$-ray flux agree with the data points considering uncertainties except the high energy end of that in region obviously under-predict the measurement for the CP model, which is similar to the result in work \citep{2012ApJ...750....3A}. It's possible that the prediction emissions from SDP model in the medium-galactic-latitude (region d) marginally lower than the observation is due to a faster diffusion with a larger halo size or due to the imperfect of parameter set. A comprehensive investigation of best-fit propagation and injection parameters, based on Bayesian inference, is left for future studies.

\bibliographystyle{aasjournal}
\bibliography{ref}

\end{document}